\documentclass[a4paper]{article}
\pdfoutput=1
\usepackage{authblk}
\usepackage{amsmath}
\usepackage{amssymb}
\usepackage{layout}
\usepackage{ifthen}
\usepackage{url}
\usepackage[pdftex,colorlinks=true,linkcolor=blue]{hyperref}
\usepackage[retainorgcmds]{IEEEtrantools}
\usepackage{graphicx}
\usepackage{xr}
\usepackage{float}

\title{A Bayesian Gamma-power-mixture survival regression model:
  predicting the recurrence of prostate cancer
  post-prostatectomy\footnote{First version deposited in arxiv; RFS
  version 1.23.1.1
.}}

\author[1]{Tommy Walker Mackay}
\author[2]{Mingtong Xu}
\author[3]{Shahrokh F. Shariat}
\author[4,5]{Roger Sewell}
\affil[1]{Mathematical Institute, University of Oxford; but at the
  time this work was done at Trinity College, Cambridge, CB2 1TQ,
  United Kingdom}
\affil[2]{Medical Research Council Laboratory of Molecular Biology, Cambridge,
  CB2 0QH, United Kingdom}
\affil[3]{Department of Urology, Medical University of Vienna, Austria}
\affil[4]{Trinity College, Cambridge, CB2 1TQ, United Kingdom}
\affil[5]{\href{mailto:roger.sewell@cantab.net}{\scriptsize{roger.sewell@cantab.net}}}
\begin{document}
\maketitle

Acknowledgements: To Monty Barlow of Cambridge Consultants for his
assistance with data access, and to Lalita Ramakrishnan for motivating
the development of the model and organising funding for Tommy Walker
Mackay and Mingtong Xu.

\begin{center}
\textbf{Abstract}
\end{center}

We constructed a Bayesian hierarchical model for predicting the
probability distribution of survival from a disease based on a number
of observed explanatory variables. This model considers an unknown
number of competing modes of death, each of which has a Gamma-power
distribution of survival time, and each of which is active with
probability given by a logistic linear regression of explanatory
variables with unknown coefficients. Choosing suitable priors we
applied this model to predicting recurrence time of prostate cancer
following radical prostatectomy based on preoperative markers,
clinical data, and operative findings, using a dataset previously
collected by Shariat et al. We estimated the apparent Shannon
information (ASI) in predictions based on the model on unseen patients
using a variety of subsets of the available data.

Results: In all the subsets examined the ASI was positive with
posterior probability greater than \mbox{0.975 .} Using only age and
results of pre-operative blood tests (PSA and biomarkers) we achieved
0.232 (0.180 to 0.290) nats ASI (0.335 (0.260 to 0.419) bits)
(posterior mean and equitailed 95\% posterior confidence
intervals). This is more than double the mean posterior ASI previously
achieved on the same dataset by a subset of the current authors using
a log-skew-Student-mixture model, and is greater than that previous
value with posterior probability greater than 0.99 . Additionally
using pre- or post-operative Gleason grades, operative findings,
clinical stage, and presence or absence of extraprostatic extension or
seminal vesicle invasion did not increase the ASI extracted. However
removing the blood-based biomarkers and replacing them with either
pre-operative Gleason grades or findings available from MRI scanning
greatly reduced the available ASI to respectively 0.077 (0.038 to
0.120) and 0.088 (0.045 to 0.132) nats (both less than the values
using blood-based biomarkers with posterior probability greater than
0.995). A greedy approach to selection of the best biomarkers gave
TGF$\beta_1$, VCAM1, IL6sR, and uPA in descending order of importance
from those examined.

We stress that these findings apply only in a set of patients who have
all undergone radical prostatectomy, and that the findings in patients
taking a watchful waiting approach might be very different.

\tableofcontents

\renewcommand{\arraystretch}{1.5}

\section{Introduction}

Prostate cancer is the most commonly diagnosed cancer and the second
leading cause of cancer death in men in the United
States\cite{Jemal}. Up to 30\% of patients experience at least
biochemical recurrence following initial therapy with curative
intent. It is believed that identifying this subset of prostate cancer
patients at the time of initial surgery would allow selection of a
subset of patients who should receive additional therapies in the
months following initial prostatectomy, thus avoiding the additional
cost, inconvenience, and morbidity associated with giving all patients
such additional treatment.  Further, it may be possible to use such
likelihood of early recurrence when choosing patient groups to enter
clinical trials for novel therapies, shortening the duration of such
trials by facilitating selection of those patients most likely to
experience early recurrence on the basis of biomarker data.  Various
authors have attempted to use combinations of biomarkers and clinical
data to predict which patients are at highest risk of recurrence.

In particular, Shariat et al\cite{Shariat} collected a dataset of 423
patients with prostate cancer and analysed it using standard Cox
proportional hazards methods\cite{Cox} and Harrell-bootstrapped
concordance\cite{Harrell}. Sewell et al\cite{Prostate} produced a
Bayesian model based on log-skew-Student mixtures, and assessed it
using apparent Shannon information\cite{ASI}, illustrating why this is
a more reliable method of assessing predictions by an algorithm than
Harrell-bootstrapped concordance. However, despite being the first
prediction method to produce statistically significantly positive ASI
from such a dataset, the amount of ASI obtained by these authors was
small (mean 0.109 nats (slightly less than one-sixth of a bit) better
than a simple exponential decay prediction that is the same for all
patients).

Now, ASI cannot exceed the true Shannon information (TSI) available in
the data about the time of relapse\cite{ASI}. However it is possible
for the ASI extracted by a bad algorithm to be less than the TSI
available, or even to be negative (i.e. more misleading than the
reference prediction), and correspondingly possible for a better
algorithm to extract more ASI. Bayesian methods extract ASI equal to
the TSI given the modelling and prior assumptions made -- but if one
model (or set of priors) reflects reality better than another, both
TSI and ASI given that model may increase over those given the less
accurate model.

In this present paper we seek to improve on the model accuracy
achieved in \cite{Prostate}, using a completely different,
Gamma-power-mixture based survival regression model, to which we again
apply Bayesian inference. In section \ref{background} we review the
background of modelling efforts, assessment of results, and previous
attempts to predict relapse of prostate cancer. In section \ref{model}
we present the current model. In section \ref{dataset} we describe the
dataset used, and give our results in detail in section \ref{results}
before discussion and conclusion in section \ref{discussion}.

\section{Background}
\label{background}

\subsection{Assessment of the results of a prediction method}
\label{assessment}

Many papers, including \cite{Shariat}, have followed
Harrell\cite{Harrell} and relied on concordance of predictions with
the order of relapse of pairs of patients whose relapse order could be
determined as a way of validating their predictions. Sewell showed in
appendix B of \cite{Prostate} how Harrell's bootstrapping can give
``82\% accuracy'' scores to algorithms that give completely random
outputs unrelated to truth; and concordance itself already relies on a
small subset of the pairs of patients, often only 7\% of the available
pairs, for whom it is possible to determine the order of relapse
(because for many pairs one will be censored before the other
relapses).

In consequence we need a better method of predicting the quality of a
prediction of time of relapse. In \cite{ASI} the Apparent Shannon
Information (ASI) in a prediction about the time of relapse is put
forward as an appropriate method, applying equally well to both
censored and uncensored patients; an appropriate method of estimating
it is also given. 

Given a reference probability distribution (such as an exponential
decay common to all patients), the ASI scores a predicted probability
distribution of relapse time by measuring the average logarithm of the
factor by which that predicted probability density exceeds the
reference probability density at the actual time of relapse of an
unseen patient (and similarly for the ratio of the predicted
probability of non-relapse by the censoring time to the reference
probability of non-relapse by the censoring time). If on average that
factor is 2 then the ASI is one bit; if it is a factor of $e$ then the
ASI is one nat (so that 1 nat = 1.44 bits, and 1 bit = 0.69
nats). (Since both predicted and reference densities have to integrate
to one, one cannot just increase the density everywhere.) On the other
hand if the prediction is misleading and on average puts the
probability density elsewhere than at the true time of relapse, so
that the predicted density at the true relapse time is below that of
the reference density, then the ASI will be negative.

We should note that any prediction method that gives purely point
predictions of relapse time, as opposed to probability distributions
on relapse time, will yield $-\infty$ on ASI, with the sole exception
that the time of relapse of every unseen patient is \textit{exactly}
at the predicted time. 

\subsection{Previous modelling of prostate cancer relapse time}
\label{previousmodelling}

Various authors have used nomograms (e.g. \cite{Kattan},
\cite{Graefen}), Cox proportional hazards models
(e.g. \cite{Shariat}), and Bayesian hierarchical models
(e.g. \cite{Prostate}); only the last has shown statistically
significantly positive ASI resulting, although a variant of a Cox
proportional hazards model where the hazard rate is assumed not to
vary with time could also do so in principle\cite{Prostate}.

However, the ASI yielded by the log-skew-Student-mixture model in
\cite{Prostate} is not even quite one sixth of a bit -- hardly
something to write home about. There are several potential
contributory reasons for this:
\begin{enumerate}
\item The explanatory variables in the dataset may not contain much
  more information about the time of relapse even given an accurate
  model;
\item The model may not accurately represent reality;
\item The priors used may not accurately represent reality;
\item In principle an arbitrary algorithm may give lower ASI than the
  available TSI given the model and priors -- however Bayesian models
  avoid this, so this is not applicable in the case of
  \cite{Prostate}.
\end{enumerate}

\section{The model}
\label{model}

We now describe the Gamma-power-mixture model used in this paper; the
way this model will be used is described in appendices 1 and 2 of
\cite{survival}; large parts of the model are similar to the survival
model used in that paper for tuberculous meningitis, though the model
of that paper did not allow for any explanatory variables. We adopt the
Bayesian paradigm and construct a generative hierarchical Bayesian
model as follows.

We suppose that there exist an unknown number $J$ of different
mechanisms causing death, and that each such mode of death has a
different lifetime distribution. Indeed throughout this description we
will use ``death'' to mean any type of event following which our
interest in the patient lapses, and in particular in this paper
including as an application example the event that there is
biochemical relapse of prostate cancer.

\subsection{Combination of different modes of death}

First let us consider a single patient $i$. Let $x$ denote a lifetime,
i.e. the time until a patient dies. Let $j\in \{1,2,...,J\}$ denote a
particular mechanism (or mode) of death. Let $x_j$ denote the time at
which mode $j$ would kill the patient; we set $x_j=\infty$ to denote
the possibility that that mode would never have killed the patient.

Then the patient's time of death is given by $$x = \min_{1\leq j\leq
  J}{x_j}.$$ In particular $x=\infty$ denotes the situation that the
patient never dies (unlikely as this is).

\subsection{Model of a single mode of death}

\label{singlemode}

We now drop the subscripts $j$, but assume that this subsection will
be repeated $J$ times with the subscript $j$s added to every random
variable, with each of the repetitions being independent as far as the
model is concerned before being conditioned on observed
data. Similarly we elide the variable $i$ indexing patients. When
later we want to refer to the complete set of $J$ values of e.g. $p$,
we will use bold face, e.g. $\mathbf{p} = (p_1,p_2,...,p_J)$ or
$\mathbf{p}_i=(p_{i,1},p_{i,2},...,p_{i,J})$ for a variable such as
$p$ which depends on both patient and mode of death.

Thus we will set $P(x|p,k,m,r)$, i.e. $P(x_j|p_j,k_j,m_j,r_j)$, to be
such that with probability $p$, $x^k$ is Gamma distributed with
parameters $m'=m$ and $r'=mr^k$, and otherwise $x=\infty$. Thus we
have $$P(x|p,k,m,r)=\left\{\begin{matrix}p|k|\frac{(mr^k)^{m}}{\Gamma(m)}x^{mk-1}e^{-m(rx)^k}
& (0 < x < \infty)\\1-p & (x=\infty)\\0 & (x \leq
0).\end{matrix}\right.$$

Here $p\in [0,1]$, $k\in \mathbb{R}\setminus\{0\}$, $m,r>0$; the
explanatory variables will enter the model through the parameter $p$.

Note that we have here a distribution which has both a discrete and a
continuous part, so that $P(x|...)$ is used as notation both for a
probability and for a probability density: in other words, we have a
continuous distribution for finite positive $x$, given by a density
function, whose interpretation is that its integral from $x_1$ to
$x_2$ is the probability that $x_1 < x < x_2$; but as we have a
non-zero probability $1-p$ that $x=\infty$, the integral from $0$
(inclusive) to $\infty$ (exclusive) of the density given by the first
line of the above formula for $P(x|p,k,m,r)$ must be $p$. On the other
hand we have a discrete distribution for $x=\infty$, and $1-p$ is a
probability, not a density.

By way of very approximate intuition: $p$ is the probability that a
particular mode of death would kill the patient at a finite time; $r$
is the reciprocal of the overall timescale to deaths of those patients
who die; $m$ governs how variable those times of death are -- the
smaller $m$ is, the more variable are the times of death; and the sign
of $k$ plays a part in determining whether the hazard rate for this
mode of death is increasing or decreasing, while the magnitude of $k$
governs how abruptly the spread of death time is cut off in the less
spread out direction. Specifically, $k = 0$ makes no sense, as then we
would have $x^k = 1$ for all $x$, and an invalid distribution would
result (so it should not be a surprise that the prior on $k$ is
bimodal with zero density at zero).

\subsection{The distribution of $p$}

In order to allow the explanatory variables to affect the survival
distributions, we now suppose that each patient $i$ has a vector
$c_i=(c_{i,1},c_{i,2},...,c_{i,V})$ of explanatory variables whose
values are known. We then set $$P(p_{i,j}|\mathbf{c}, \mathbf{\beta})
= \frac{1}{1+e^{l_{i,j}}}$$ where $$l_{i,j}=
\sum_{v=1}^V{c_{i,v}\beta_{v,j}},$$ where $\beta_{v,j}$ is the
coefficient of variable $c_v$ for mode of death $j$, to be determined
by inference from the dataset.

\subsection{Priors on the parameters}

\label{priors}

We specify the priors on the parameters in two stages. First, we
specify their general form, and second we choose specific values for
the hyperparameters that then specify a unique prior.

\subsubsection{General form of the priors}

\label{parmpriors}

The total number $J$ of modes is itself to be considered a random
variable, on which we put the prior $$P(J|\alpha_\text{J})=
(1-\alpha_\text{J})\alpha_\text{J}^{J-1}$$ for $J\in \mathbb{N}^* =
\{1,2,...\}$ and for some fixed $\alpha_\text{J}\in [0,1).$ 

The prior for the parameters $m,r,k, \beta$ of each mode of death are taken
to be independent, and as follows.

We take the prior on $r$ to be Gamma, with parameters
$m_\text{r},r_\text{r}>0$, so
that $$P(r|m_\text{r},r_\text{r})=\frac{r_\text{r}^{m_\text{r}}}{\Gamma(m_\text{r})}
r^{m_\text{r}-1}e^{-r_\text{r}r}.$$

We take the prior on each of the parameters $k$ and $m$ to be the
conjugate prior on each with respect to this parameterisation. Thus
for positive real parameters $a_\text{m},b_\text{m}$ we
have $$P(m|a_\text{m},b_\text{m}) \propto \frac{m^{b_\text{m} m}
  e^{-(a_\text{m} + b_\text{m})m}}{\Gamma(m) ^ {b_\text{m}}}.$$
Similarly for parameters $N_\text{k}\in\mathbb{N}$,
$a_\text{k}\in\mathbb{R}_+$ and
$\mathbf{b}_\text{k}=(b_1,b_2,...,b_{N_\text{k}}),\mathbf{c}_\text{k}=(c_1,c_2,...,c_{N_\text{k}})
\in\mathbb{R}_+^{N_\text{k}}$ we
have $$P(k|a_\text{k},\mathbf{b}_\text{k},\mathbf{c}_\text{k})\propto
|k|^{a_\text{k}} \prod_{n=1}^{N_\text{k}}{b_n^{kc_n}e^{-c_n b_n^k}}.$$

In the case of $\beta_{v,j}$ we
set $$P(\beta_{v,j}|\gamma)=\frac{e^{\gamma\beta_{v,j}}}{(1 +
  e^{\gamma\beta_{v,j}})^2}.$$

\subsubsection{Specific values of the hyperparameters and the
  resulting priors}

The priors were chosen to be uninformative and very wide, with the
exception of the prior on $\beta_{v,j}$, for which setting it too wide
turned out to have the undesirable effect of implying a high
probability that each mode of death would either be always active for
all patients or always inactive. Specifically we took the following
values:

The specific parameter values chosen were as
follows: $$a_\text{J}=0.8$$ $$\gamma=1$$ (where we assume that all
explanatory variables have been shifted to have mean zero then scaled
to variance one over the
database, with the exception of a single constant variable equal to
one for all patients) $$a_\text{m}=b_\text{m}=1$$ $$m_\text{r}=0.5$$
 $$r_\text{r}=30 \text{ days}$$ $$N_\text{k}=2$$ $$a_\text{k}=1$$ 
$$\mathbf{b}_\text{k}=(0.2,0.2)$$ $$\mathbf{c}_\text{k}=(0.5,0.5).$$

\begin{figure}
\begin{center}

\includegraphics[scale=0.4]{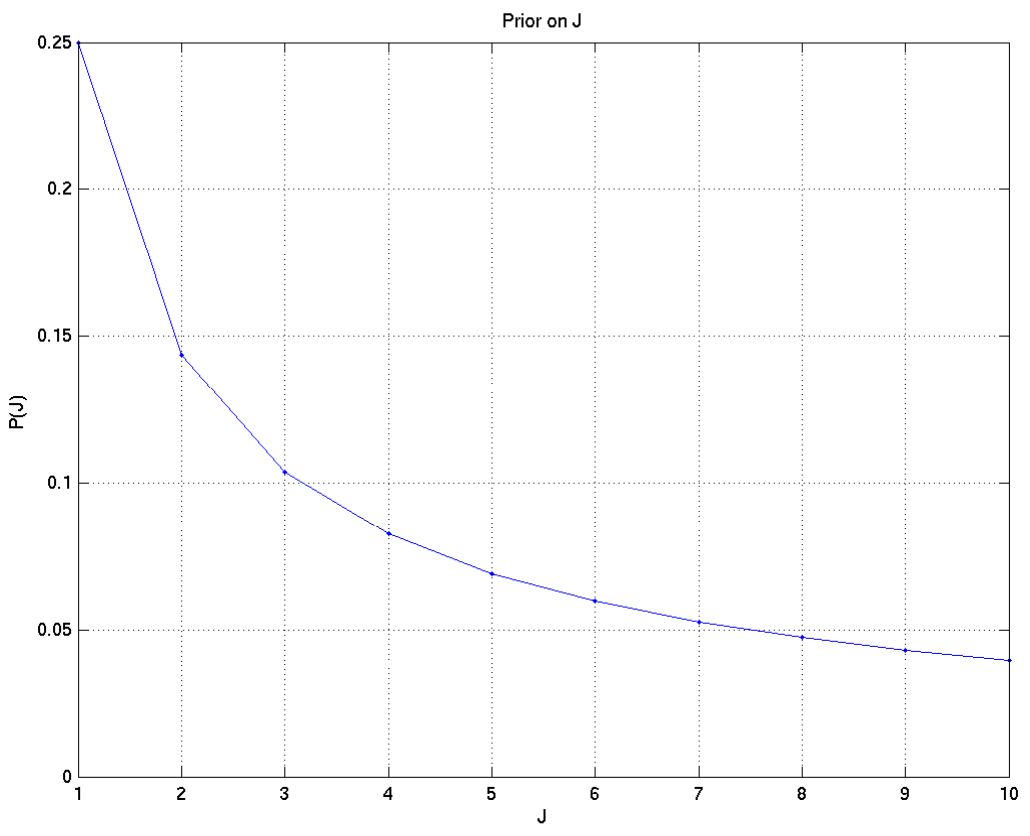}

\end{center}

\caption{Prior on $J$, the number of different modes of death.
\label{Jprior}
}

\end{figure}

\begin{figure}
\begin{center}

\includegraphics[scale=0.4]{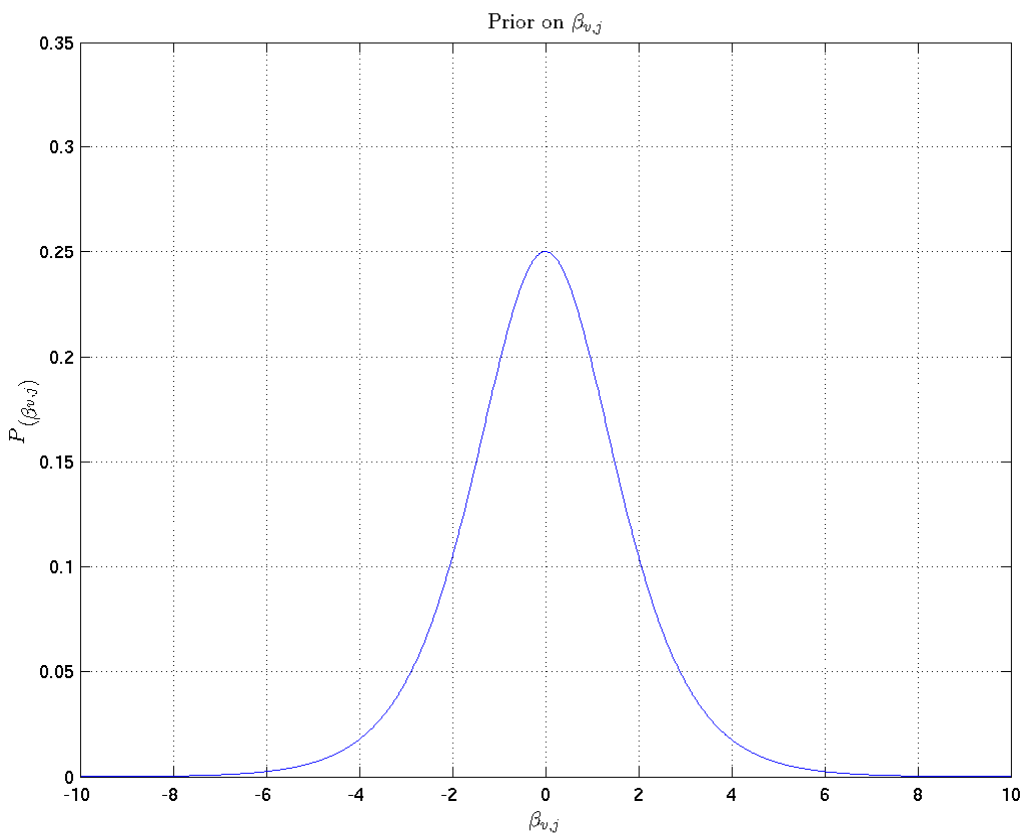}

\end{center}

\caption{Prior on $\beta_{v,j}$, the coefficient of $c_{i,v}$ for mode
  $j$ of death.
\label{betaprior}
}

\end{figure}

\begin{figure}
\begin{center}

\includegraphics[scale=0.4]{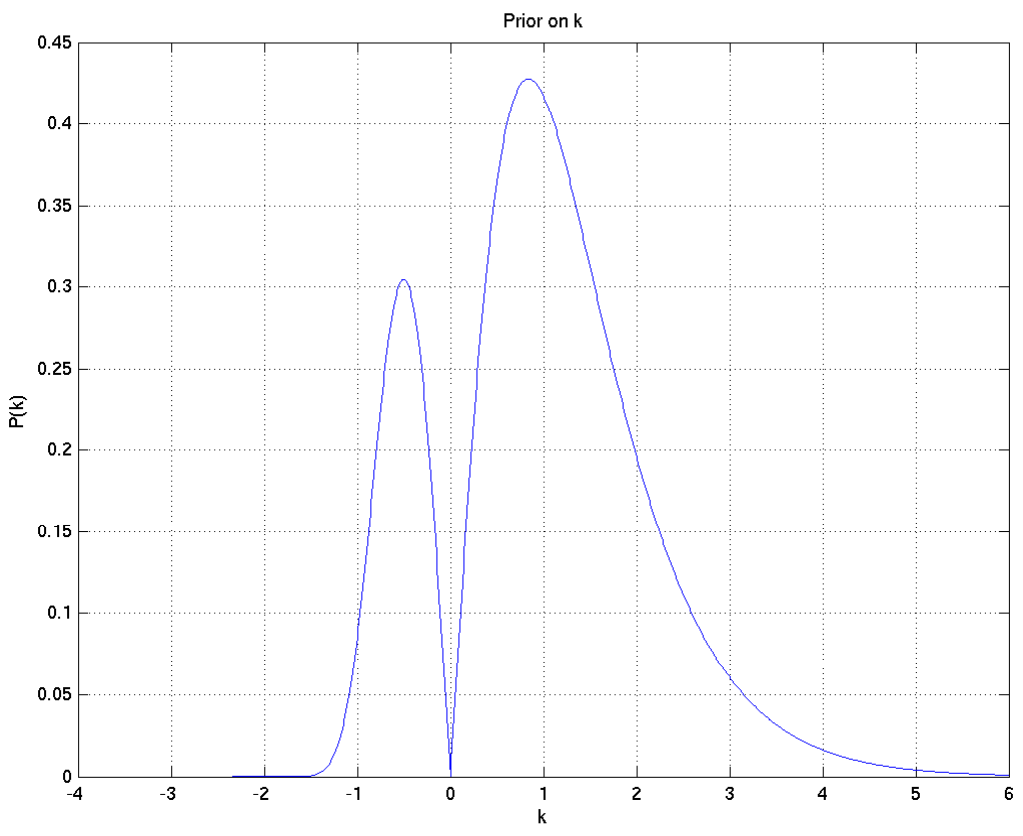}

\end{center}

\caption{Prior on $k$.
\label{kprior}
}

\end{figure}

\begin{figure}
\begin{center}

\includegraphics[scale=0.4]{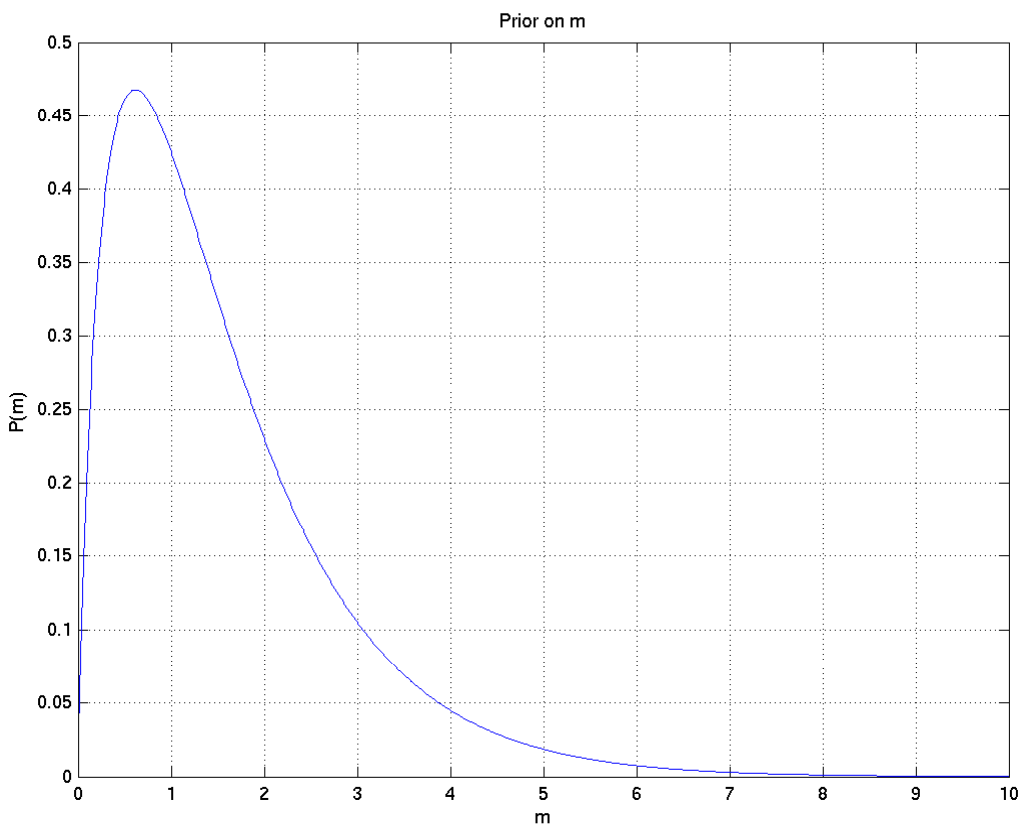}

\end{center}

\caption{Prior on $m$.
\label{mprior}
}

\end{figure}

\begin{figure}
\begin{center}

\includegraphics[scale=0.4]{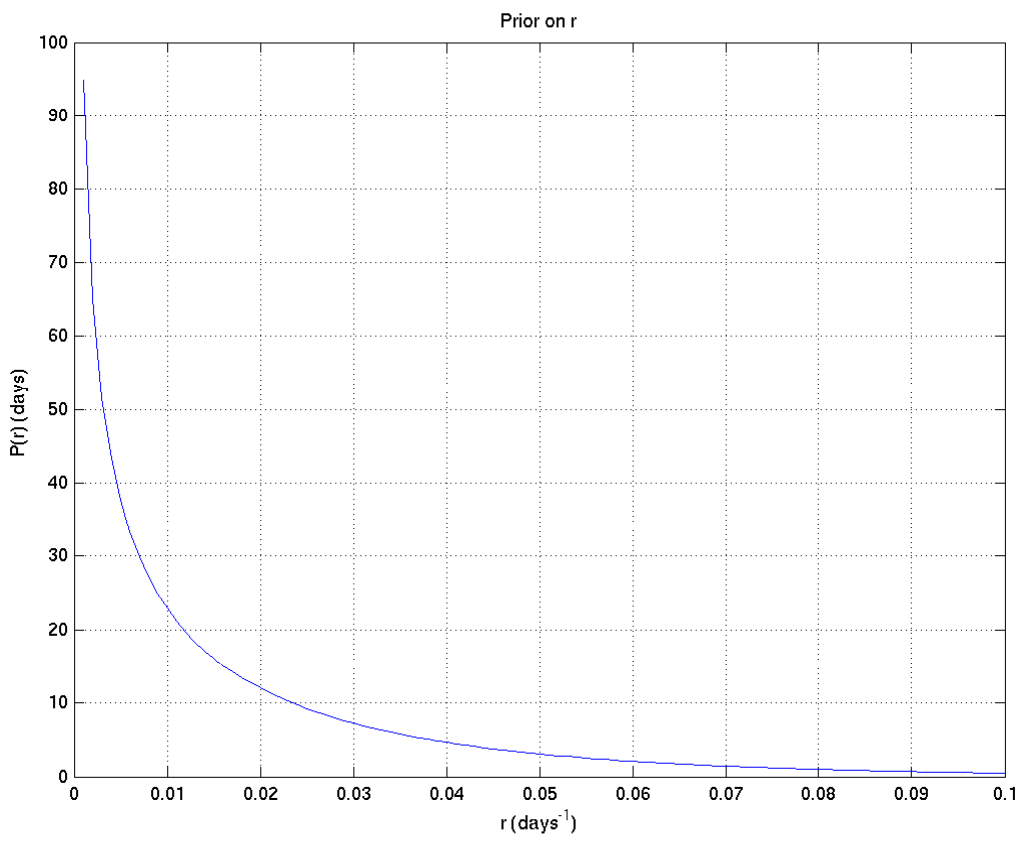}

\end{center}

\caption{Prior on $r$.
\label{rprior}
}

\end{figure}

\begin{figure}
\begin{center}

\includegraphics[scale=0.4]{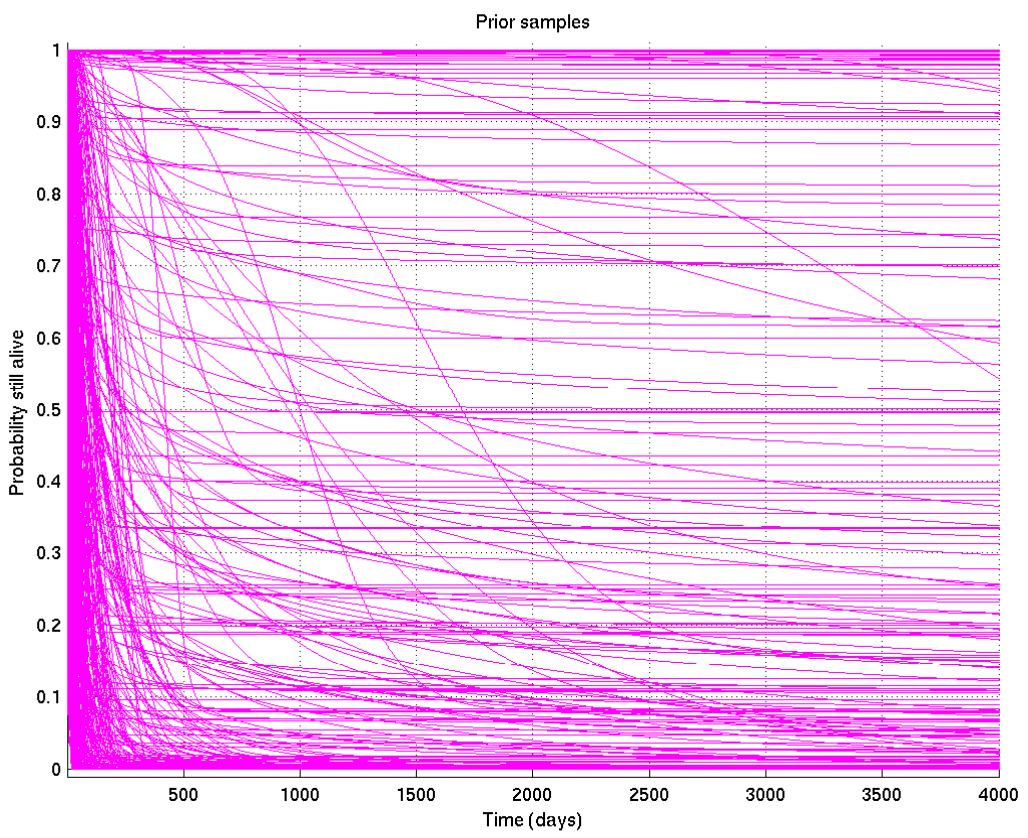}

\end{center}

\caption{Samples from resulting prior on survival probability against
  time. Here the variables $c_{i,v}$ have been taken from different random
  patients in the dataset for each curve, while the $\beta_{v,j}$ and
  other variables used are random samples from the prior.
\label{survivalpriorsamples}
}

\end{figure}

\begin{figure}
\begin{center}

\includegraphics[scale=0.4]{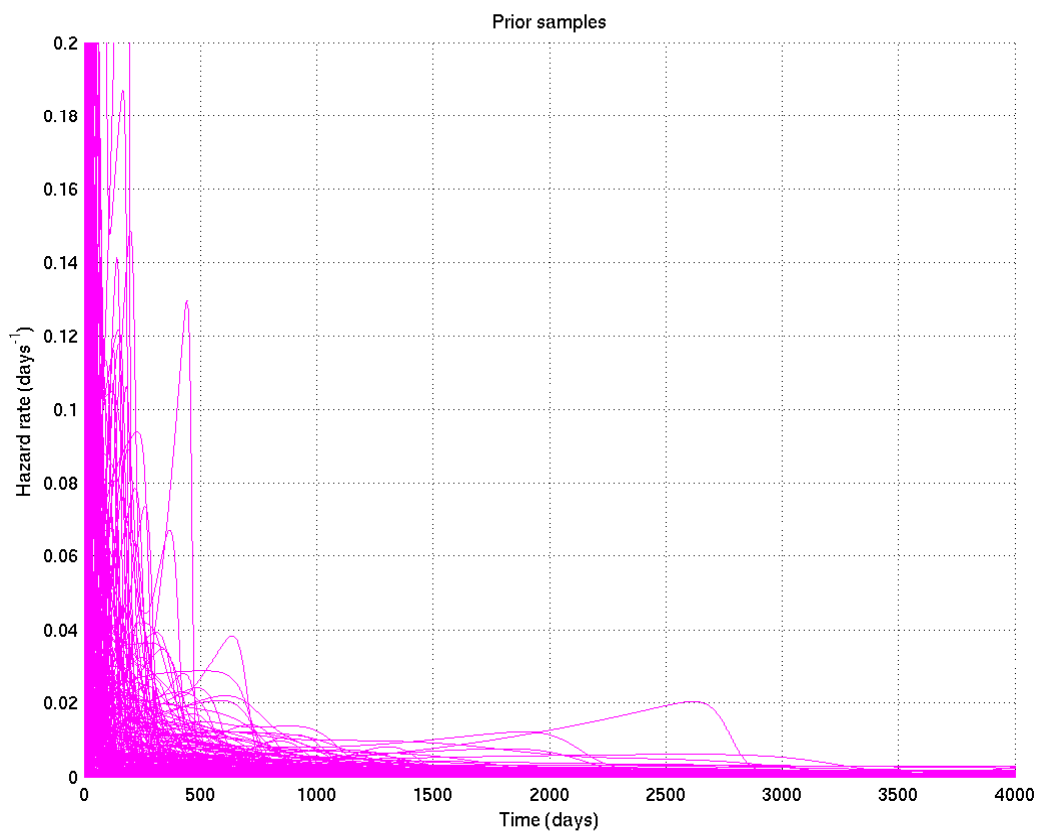}

\end{center}

\caption{Samples from resulting prior on hazard rate against
  time. Here the variables $c_{i,v}$ have been taken from different
  random patients in the dataset for each curve, while the
  $\beta_{v,j}$ and other variables used are random samples from the
  prior.
\label{hazardpriorsamples}
}

\end{figure}

\begin{figure}
\begin{center}

\includegraphics[scale=0.4]{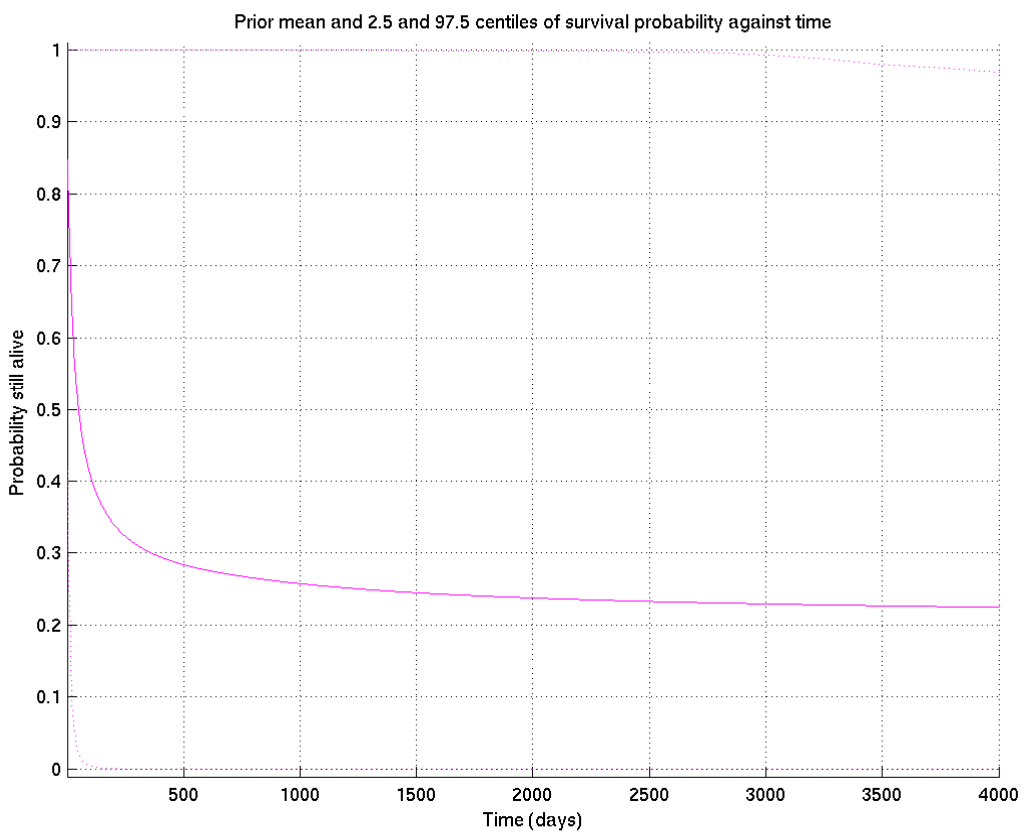}

\end{center}

\caption{Mean and 2.5\% and 97.5\% centiles of prior on survival
  probability against time.
\label{survivalpriormean}
}

\end{figure}

\begin{figure}
\begin{center}

\includegraphics[scale=0.4]{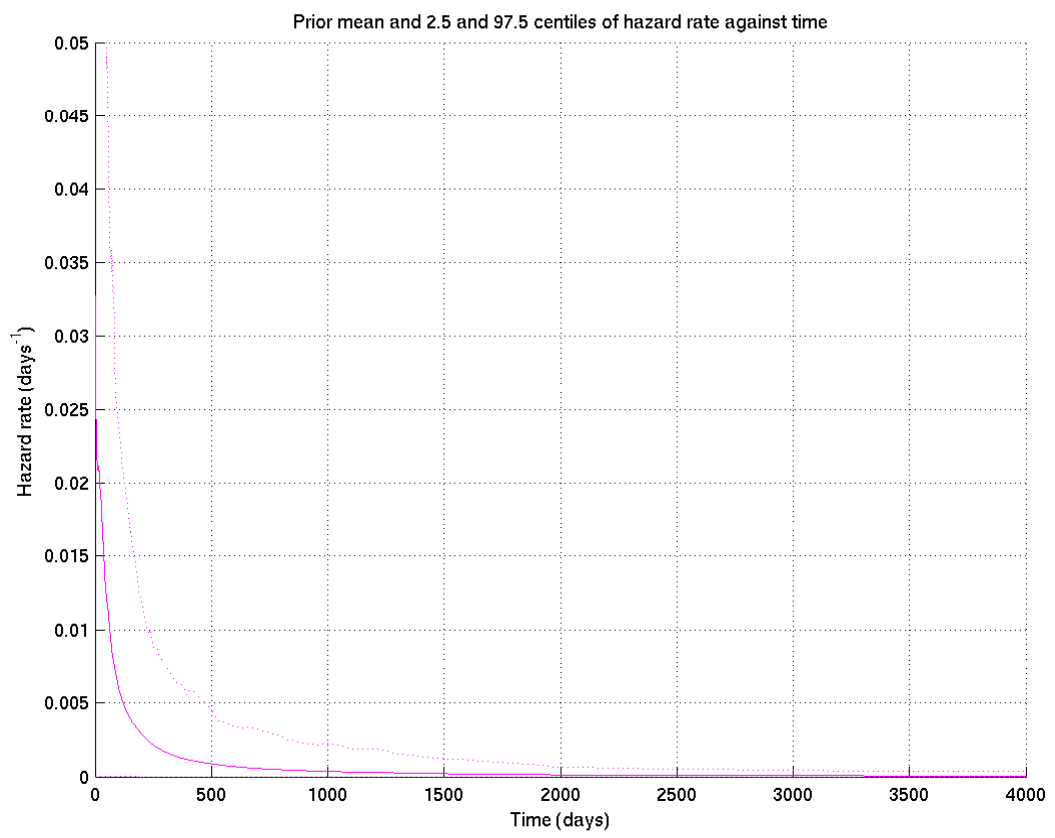}

\end{center}

\caption{Mean and 2.5\% and 97.5\% centiles of prior on hazard rate against time.
\label{hazardpriormean}
}

\end{figure}

These result in the following depicted distributions for
$J,\beta_{v,j},m_j,r_j,k_j$, and hence for the depicted samples from the
distributions for survival probability and hazard rate against time as
well as the mean and 2.5\% and 97.5\% centiles for the last two: see
Figures \ref{Jprior} to \ref{hazardpriormean}.

\subsection{MCMC methodology}

We introduce additional variables $j_i$ for each patient $i$ which
indicate whether the time of death was censored (value 0) or was
caused by a particular mode $j$ of death (value $j\neq 0$ unknown). We
also introduce variables $x_{i,j}$ of unknown values giving for each
patient the time of death that would have resulted from mode $j$ if no
other modes had killed the patient first. These variables take the
specific value $x_{i,j}=\infty$ if mode $j$ would in fact not have
killed patient $i$ at any finite time.

We initialise the parameters
$J,\mathbf{\beta},\mathbf{m},\mathbf{r},\mathbf{k}$ from the prior and
initialise the additional variables $\mathbf{j}$ and $\mathbf{x}$
randomly to any set compatible with those and the observed variables
$\hat{\mathbf{x}}$. These variables then form $\theta_1 =
(J,\mathbf{\beta},\mathbf{m},\mathbf{r},\mathbf{k},\mathbf{j},\mathbf{x})$,
the first of a sequence of samples $(\theta_n)_{n=1,...}$ to be drawn.

\subsection{Sampling methods}

A thorough review of the methods underlying all of the following is
available in either \cite{rnealMCMC} or in \cite{Dagpunar} except
where otherwise indicated.

The key point is that if we resample each variable by a method that
satisfies detailed balance, and given other weak conditions which are
here fulfilled, Feller's theorem \cite{rnealMCMC} then guarantees the
the sequence of samples $(\theta_n)$ will eventually converge to a
sequence of samples from the desired distribution
$P(\theta|\hat{\mathbf{x}})$. The samples in this sequence will not be
independent of each other, though the conditional distribution of
$\theta_{n_1}$ given $\theta_{n_0}$ will also converge to
$P(\theta|\hat{\mathbf{x}})$ as $n_1\to \infty$ with $n_0$ fixed,
i.e. to independence.

Sampling from the posterior was done by the MCMC technique of Gibbs
sampling, i.e. sampling from the following distributions
palindromically:
\begin{enumerate}

\item
  $P(\mathbf{k}|\hat{\mathbf{x}},\mathbf{x},J,\mathbf{j},\mathbf{m},\mathbf{r},\mathbf{\beta})$. This
  distribution has two parts ($k_j>0$ and $k_j<0$), each of which is
  log-concave. We therefore first resample the sign of each $k_j$
  using the Metropolis-Hastings algorithm \cite{rnealMCMC}, then use
  adaptive rejection sampling \cite{ARSwikipedia} to resample the
  magnitude of $k_j$ given its sign, then resample the sign again to
  maintain detailed balance.

\item
  $P(\mathbf{m}|\hat{\mathbf{x}},\mathbf{x},J,\mathbf{j},\mathbf{r},\mathbf{k},\mathbf{\beta})$. This
  distribution is log concave, so we may use adaptive rejection
  \cite{ARSwikipedia} sampling to sample from it.

\item
  $P(\mathbf{r}|\hat{\mathbf{x}},\mathbf{x},J,\mathbf{j},\mathbf{m},\mathbf{k},\mathbf{\beta})$. For
  each $j$, this distribution is in general a product of a Gamma
  distribution on $r_j$ and a much narrower Gamma distribution on
  $r_j^{k_j}$. We therefore sample from the Gamma relevant to the
  latter \cite{Dagpunar}, using this as a proposal distribution for
  the Metropolis-Hastings algorithm \cite{rnealMCMC}, resulting in the
  Hastings ratio coming from the Gamma on $r_j$.

\item
  $P(\mathbf{j}|\hat{\mathbf{x}},J,\mathbf{m},\mathbf{r},\mathbf{k},\mathbf{\beta})$
  then
  $P(\mathbf{x}|\hat{\mathbf{x}},J,\mathbf{j},\mathbf{m},\mathbf{r},\mathbf{k},\mathbf{\beta})$. The
  first of these is a discrete distribution which is trivial to sample
  from, and the second reduces to a truncated Gamma distribution. To
  sample from the latter we divide into two cases: if the shape
  parameter is $\geq 1$ the distribution is log-concave and we can use
  adaptive rejection sampling \cite{ARSwikipedia}; otherwise we use
  Metropolis-Hastings \cite{rnealMCMC} with either an exponential or a
  Gamma proposal distribution, depending which is estimated to be
  likely to be quicker given the other parameters.

\item
  $P(J|\hat{\mathbf{x}},\mathbf{x},\mathbf{j},\mathbf{m},\mathbf{r},\mathbf{k},\mathbf{\beta})$
  (where only values of $j$ unused in $\mathbf{j}$ are allowed to be
  removed) followed, if $J$ has increased, by sampling the new
  elements of $\mathbf{m},\mathbf{r},\mathbf{k},\mathbf{p}$ from the
  prior distributions on these variables. Resampling of $J$ uses a
  discrete conditional distribution, and is done using a proposal to
  either increase or decrease $J$ by 1, and applying the appropriate
  Hastings ratio \cite{rnealMCMC} to reject the proposal in such a way
  as to achieve detailed balance.

\item
  $P(\mathbf{\beta}|\hat{\mathbf{x}},\mathbf{x},J,\mathbf{j},\mathbf{m},\mathbf{r},\mathbf{k})$. In
  this case the vector $(\beta_{v,1})_{v=1,...,V}$ is independent of
  the $\beta_{v,j}$ for $j\neq 1$, so we resample each such vector
  $\beta=(\beta_{v,j})_{v=1,...,V}$ separately for $j=1,...,J$. The
  conditional distribution we need to sample from has the
  form $$P(\beta|\mathbf{X}) \propto \prod_{g=1}^G{\frac{1}{1 +
      e^{X_{g,:}\beta}}}$$ for some matrix $\mathbf{X}$ of size
  $G\times V$ with $\mathbf{X}_{g,:}$ denoting row $g$ of the
  matrix. The following resampling method is the result of trying a
  range of alternatives, and choosing that which maximises mobility.

  To maintain numerical stability and ensure good resampling mobility,
  we first change the coordinate system by rescaling $\mathbf{X}$,
  replacing it with $$\mathbf{Y} = \mathbf{X}\mathbf{C}^{-1}$$ where
  $\mathbf{C}$, the Cholesky decomposition of the symmetrised version
  of $\mathbf{X}$, is an upper triangular matrix such that
  $\mathbf{C}'\mathbf{C}=\mathbf{X'X}$; this has the effect of
  changing the scale of the columns of $\mathbf{X}$ to be one, i.e. of
  making the eigenvalues of $\mathbf{Y}'\mathbf{Y}$ to be one (and
  indeed $\mathbf{Y}'\mathbf{Y}$ to be the identity). We then apply a
  rotation matrix $\mathbf{Q}$ to $\mathbf{Y}$ on the right, replacing
  $\mathbf{Y}$ with $\mathbf{YQ}$; here $\mathbf{Q}$ is randomly
  chosen from the uniform distribution on all rotation matrices by
  taking it to be the rotational (orthogonal) component of the
  QR-decomposition of a matrix all of whose entries are drawn from
  independent unit Gaussians. We apply the corresponding
  transformations to the current value of $\beta$, so that the
  distribution of the transformed $\beta$ using $\mathbf{Y}$
  corresponds to that of the original $\beta$ using $\mathbf{X}$.

  We then apply one step of Newton's method, starting from the current
  point, using the derivative and Hessian $\mathbf{H}$ of the log
  density, to estimate the mode of the conditional distribution from
  which we wish to sample; we start our proposal by moving to
  $\mathbf{\hat{\beta}}$, half way to that estimated mode from the
  current value of $\beta$. (We also add $10^{-4}\mathbf{I}$ times the
  largest singular value of $\mathbf{Y}$ to $\mathbf{H}$ before
  inverting it to avoid awkward singularities arising.)

  We complete our proposal distribution by proposing a point that is
  Student distributed with shape $m=2$ and variance $-\mathbf{H}/2$
  about $\mathbf{\hat{\beta}}$. We then apply the Hastings ratio (see
  \cite{rnealMCMC}) to determine whether to accept newly proposed
  value of $\beta$ or to reissue the previous one, before finally
  transforming the result back to the original coordinate system.

\end{enumerate}

We apply simulated annealing\cite{rnealMCMC} over the first 1000 MCMC
samples, then continue at coolness one for a further 7000 samples,
discarding the first 2050 of the resulting 8000 samples. We anneal
from a distribution $P_0(\mathbf{x})$ at coolness zero to the modelled
distribution $P(\mathbf{x}|\theta)$ at coolness one, setting the
distribution at coolness $t$ to be $$P_t(\mathbf{x}|\theta) =
P_0(\mathbf{x})^{1-t}P(\mathbf{x}|\theta)^t,$$ where $t$ is the
coolness and $P_0(\mathbf{x})$, independent for each patient and mode
of death, is an equiprobable mixture of a zero-centred Cauchy of width
30 days truncated below at zero and a point distribution at
infinity. 

We set the coolness $t$ by requiring $\log\left(\frac{t}{1-t}\right)$
to be linearly spaced over the 1000 samples for which simulated
annealing is in force.

While this suffices to define the annealing scheme, the
details of the modified resampling are non-trivial.

Convergence was checked by using synthetic data and ensuring that the
truth for survival and hazard rate lies within the equitailed 95\%
posterior confidence interval at least 95\% of the time.

\subsection{Estimation of the ASI}

\begin{figure}
\begin{center}

\includegraphics[scale=0.4]{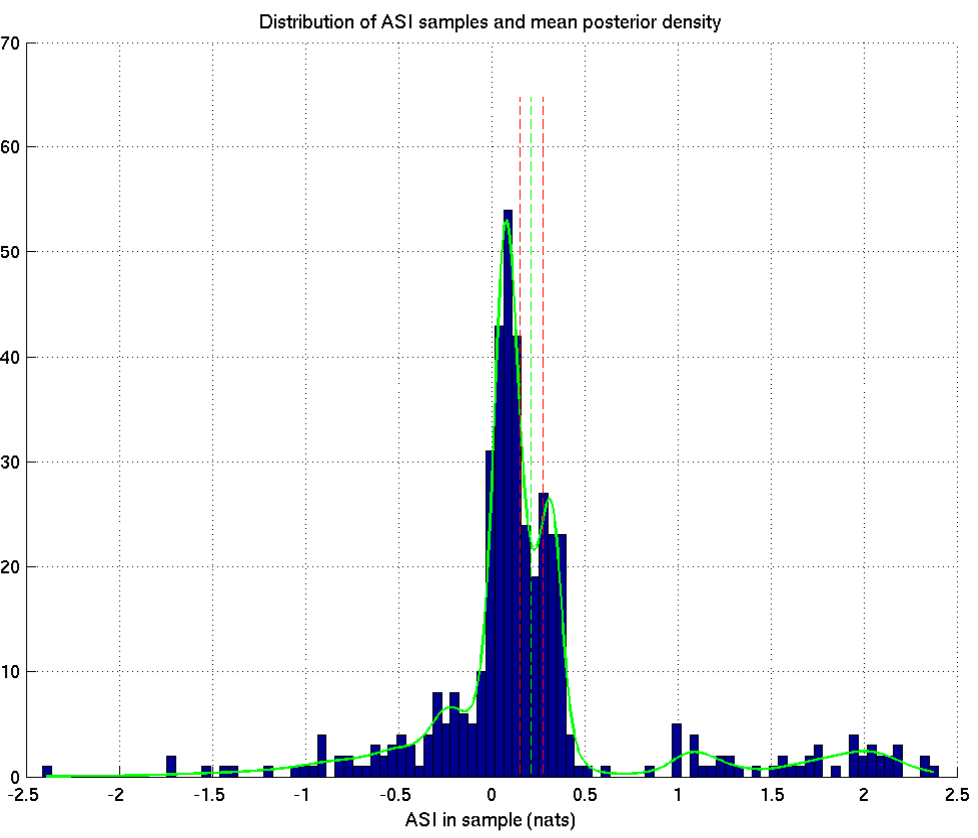}

\end{center}

\caption{Samples of the ASI measured on individual patients using all
  the available explanatory variables (dark blue histogram), with mean
  posterior skew-Student-mixture (green curve), mean posterior ASI
  (green dashed line) and equitailed 95\% posterior confidence
  interval for the mean ASI (red dashed lines).
\label{ASIsamples}
}

\end{figure}

\begin{figure}
\begin{center}

\includegraphics[scale=0.4]{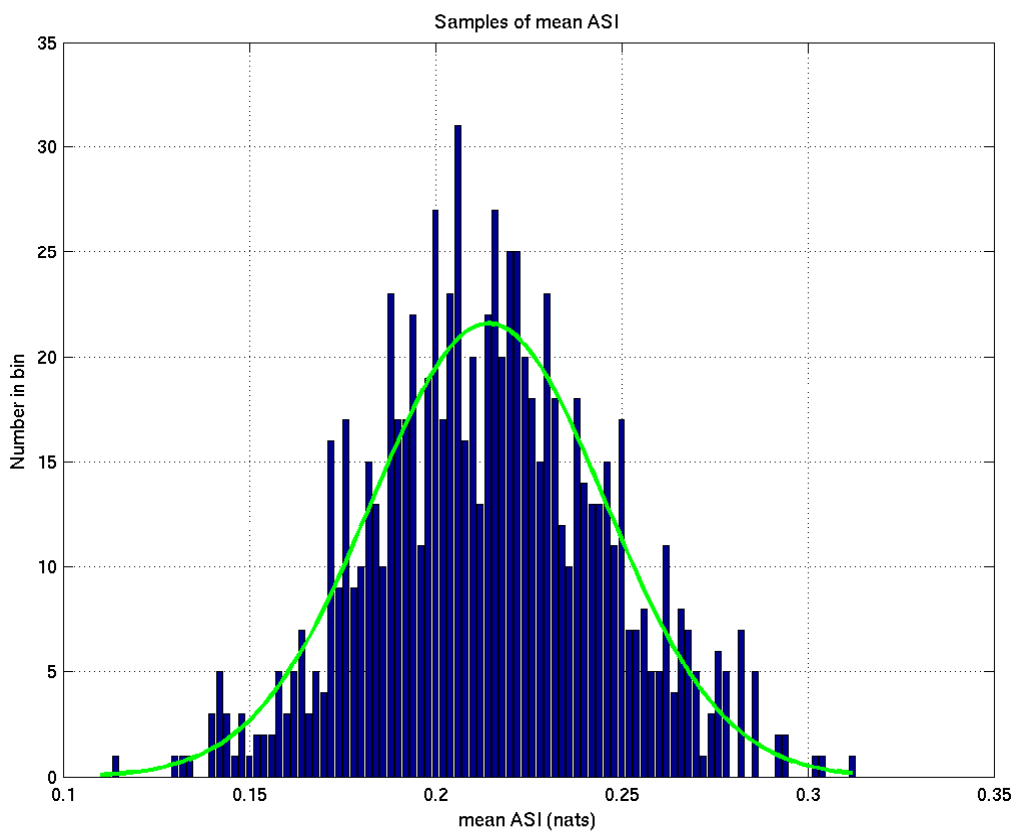}

\end{center}

\caption{Posterior samples of the mean ASI taken from posterior
  samples of the skew-Student mixture inferred from the histogram
  of figure \ref{ASIsamples}, with the Gaussian of mean and variance
  corresponding to the empirical values from these samples.
\label{ASImeansamples}
}

\end{figure}

For each subset of the explanatory variables considered, training was
done on a randomly selected half of the patients, and ASI estimated on
the other half, then swapping the roles of the two halves, exactly as
in the 2/2 scenario of \cite{ASI}. Having measured the individual
samples of ASI on each patient, getting a histogram such as that of
Figure \ref{ASIsamples}, we used a skew-Student model identical to
that used in \cite{ASI} to estimate the mean ASI; it gave us a set of
posterior samples of the mean ASI such as that shown in Figure
\ref{ASImeansamples} along with its Gaussian fit according to the
empirical mean and variance of the sample set.

For statistical comparison of mean ASI from one subset of the
explanatory variables with that from another, the probability of one
being greater than the other was estimated on the basis of 850 samples
of mean ASI from each. Where one of the comparands was from previously
published work whose samples are no longer available, we both applied
Gaussian assumptions (reasonable on the basis e.g. of Figure
\ref{ASImeansamples}), and also separately determined bounds that
could be proven from the published mean and centiles.

\section{The dataset}
\label{dataset}

The dataset has previously been described in \cite{Shariat}. Briefly,
pre-operative biomarkers PSA, free PSA, TGF$\beta_1$, IL6sR, IL6,
VCAM1, VEGF, endoglin, PAI1, uPA, and uPAR were measured in
peripheral blood on 423 patients with biopsy-proven localised prostate
cancer, all of whom underwent radical prostatectomy and bilateral
lymphadenectomy. Also recorded were the major and minor Gleason grades
from pre-operative biopsy and from operative samples, age, and whether
or not there was seminal vesicle invasion, extraprostatic extension,
tumour at the surgical margins, or metastases in lymph nodes.

Patients were considered to have relapsed when PSA rose above 0.2
$\mu$g/litre and remained so on a confirmation reading.

Patients were followed for an average of 38.9 months, by which time
they had either relapsed or were considered censored when follow-up
ceased. 

\section{Results}
\label{results}

The ASI obtained in predictions of relapse time (or of non-relapse at
censoring time) was as shown in Table \ref{resultstable} for a variety
of subsets of the explanatory variables. That obtained using the full
set of explanatory variables was approximately double that obtained in
\cite{Prostate} using the same set of variables, and was greater with
probability at least 0.859 without distributional assumptions or at
least 0.975 using the Gaussian assumption supported e.g. by Figure
\ref{ASImeansamples}. The values obtained using any subset that
included the biomarkers were greater than those using only age, PSA,
MRI-discoverable variables, pre-op Gleason grades, and surgically
discoverable variables with probability greater than 0.99 (Table
\ref{comparisontable}).

\begin{table}[ht]
\begin{center}
\begin{tabular}{rrrr}
&\multicolumn{3}{c}{\textbf{(nats)}}\\
\textbf{Variables} & \textbf{2.5\% ASI} & \textbf{mean ASI} & \textbf{97.5\% ASI}\\
Prev &   0.024 &   0.109 &   0.195 \\
A &   0.015 &   0.047 &   0.078 \\
AG &   0.038 &   0.077 &   0.120 \\
AM &   0.045 &   0.088 &   0.132 \\
AMS &   0.074 &   0.124 &   0.178 \\
AMGS &   0.042 &   0.105 &   0.158 \\
AB &   0.180 &   0.232 &   0.290 \\
ABM &   0.171 &   0.224 &   0.279 \\
ABGM &   0.154 &   0.207 &   0.267 \\
ABGMS &   0.135 &   0.197 &   0.256 \\
\end{tabular}
\caption{ASI obtained from each subset of the variables:\\
A = age, PSA.\\
G = pre-op Gleason grades.\\
M = extraprostatic extension, seminal vesicle invasion, clinical stage
(MRI discoverable variables).\\
B = biomarkers in peripheral blood (free PSA, TGF-h1, sIL-6R, IL-6,
VCAM-1, VEGF, endoglin,\\ \mbox{\hspace{1 em}}\hspace{\stretch{1}} PAI-1, uPA, and uPAR).\\
S = tumour presence in operative margin, or in lymph nodes, and
Gleason grades \\ \mbox{\hspace{1 em}} \hspace{\stretch{1}} in surgically removed tissue (surgically discoverable
variables).\\
Prev = result obtained in \cite{Prostate} using all variables.\\
\label{resultstable}
}
\end{center}
\end{table}

\begin{table}[ht]
\begin{center}
\begin{tabular}{rrrrrrrrrrr}
&\multicolumn{10}{c}{\textbf{Variables B}}\\
\textbf{Variables A} & Prev & A & AG & AM & AMS & AMGS & AB & ABM & ABGM & ABGMS \\
Prev &   0.500 &   0.908 &   0.747 &   0.669 &   0.384 &   0.526 & \textbf{  0.009} & \textbf{  0.014} & \textbf{  0.029} & \textbf{  0.050} \\
A &   0.092 &   0.500 &   0.129 &   0.078 & \textbf{  0.009} & \textbf{  0.047} & \textbf{  0.000} & \textbf{  0.000} & \textbf{  0.000} & \textbf{  0.000} \\
AG &   0.253 &   0.871 &   0.500 &   0.370 &   0.089 &   0.213 & \textbf{  0.000} & \textbf{  0.000} & \textbf{  0.000} & \textbf{  0.001} \\
AM &   0.331 &   0.922 &   0.630 &   0.500 &   0.159 &   0.311 & \textbf{  0.000} & \textbf{  0.000} & \textbf{  0.000} & \textbf{  0.003} \\
AMS &   0.616 & \textbf{  0.991} &   0.911 &   0.841 &   0.500 &   0.676 & \textbf{  0.003} & \textbf{  0.006} & \textbf{  0.015} & \textbf{  0.039} \\
AMGS &   0.474 & \textbf{  0.953} &   0.787 &   0.689 &   0.324 &   0.500 & \textbf{  0.001} & \textbf{  0.001} & \textbf{  0.004} & \textbf{  0.014} \\
AB & \textbf{  0.991} & \textbf{  1.000} & \textbf{  1.000} & \textbf{  1.000} & \textbf{  0.997} & \textbf{  0.999} &   0.500 &   0.583 &   0.743 &   0.797 \\
ABM & \textbf{  0.986} & \textbf{  1.000} & \textbf{  1.000} & \textbf{  1.000} & \textbf{  0.994} & \textbf{  0.999} &   0.417 &   0.500 &   0.674 &   0.737 \\
ABGM & \textbf{  0.971} & \textbf{  1.000} & \textbf{  1.000} & \textbf{  1.000} & \textbf{  0.985} & \textbf{  0.996} &   0.257 &   0.326 &   0.500 &   0.582 \\
ABGMS & \textbf{  0.950} & \textbf{  1.000} & \textbf{  0.999} & \textbf{  0.997} & \textbf{  0.961} & \textbf{  0.986} &   0.203 &   0.263 &   0.418 &   0.500 \\
\end{tabular}
\caption{Probability that ASI from variable subset A is greater than that from variable subset B.\\
A = age, PSA.\\
G = pre-op Gleason grades.\\
M = extraprostatic extension, seminal vesicle invasion, clinical stage
(MRI discoverable variables).\\
B = biomarkers in peripheral blood (free PSA, TGF-h1, sIL-6R, IL-6,
VCAM-1, VEGF, endoglin,\\ \mbox{\hspace{1 em}}\hspace{\stretch{1}} PAI-1, uPA, and uPAR).\\
S = tumour presence in operative margin, or in lymph nodes, and
Gleason grades \\ \mbox{\hspace{1 em}} \hspace{\stretch{1}} in surgically removed tissue (surgically discoverable
variables).\\
Prev = result obtained in \cite{Prostate} using all variables.\\
\label{comparisontable}
}
\end{center}
\end{table}

\begin{figure}
\begin{center}

\includegraphics[scale=0.8]{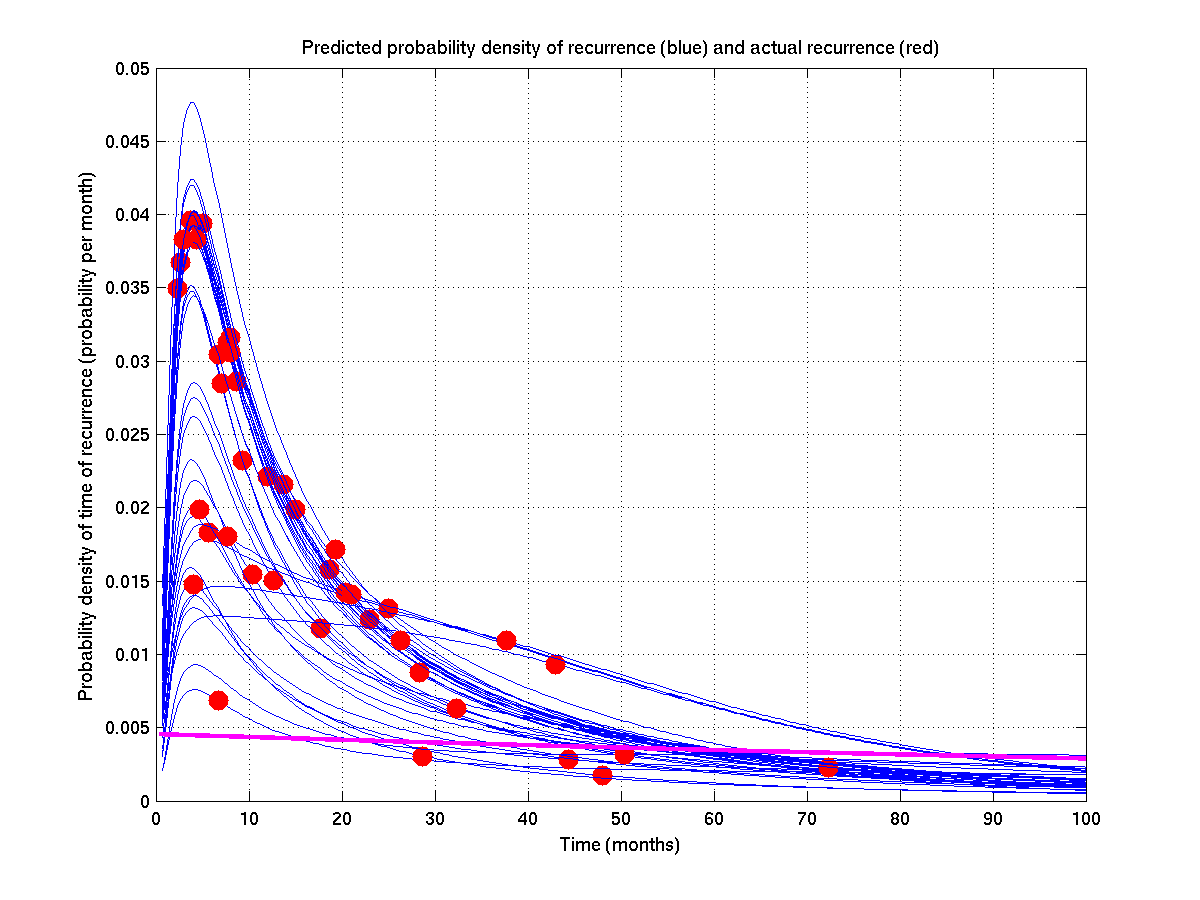}

\end{center}

\caption{Illustration of the predictions made by the model using all
  the potential explanatory variables. The magenta curve is the
  density function of the reference prior for measuring the ASI. The
  blue curves are predictions of the probability density of time of
  relapse for individual patients, unseen during training, who did in
  fact relapse; note that they integrate to $\leq 1$ with the
  remaining probability being a point mass at $\infty$. The red blobs
  indicate the time of actual relapse. The sample of the ASI from an
  individual such patient is the logarithm of (the height of the red
  blob divided by the height of the magenta line at the same time
  point).
\label{diedplots}
}

\end{figure}

\begin{figure}
\begin{center}

\includegraphics[scale=0.8]{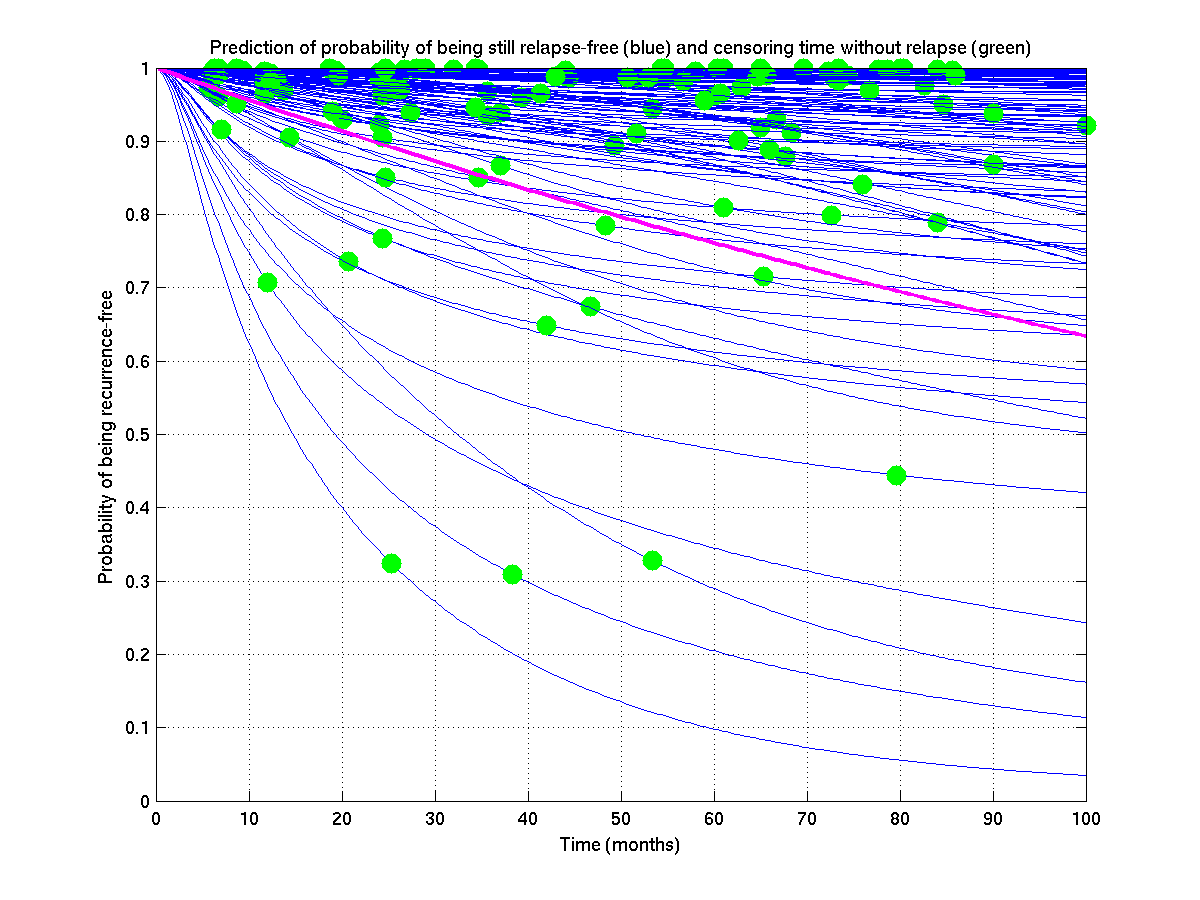}

\end{center}

\caption{Illustration of the predictions made by the model using all
  the potential explanatory variables. The magenta curve is the
  decumulative probability function of the reference prior for
  measuring the ASI. The blue curves are predictions of the
  probability density of time of relapse for individual patients,
  unseen during training, who did not in fact relapse before being
  censored. The green blobs indicate the time of actual censoring. The
  sample of the ASI from an individual such patient is the logarithm
  of (the height of the green blob divided by the height of the
  magenta line at the same time point).
\label{liveplots}
}

\end{figure}

The calculation of the ASI is illustrated intuitively in figures
\ref{diedplots} and \ref{liveplots}. Where a red (resp. green) blob is
above the magenta reference prior the patient contributes a positive
sample of the ASI; where below, a negative sample; both contribute to
the histogram of figure \ref{ASIsamples}.

Taking a greedy approach to determine which particular biomarkers
would provide how much information gave the incremental findings in
Table \ref{biomarkerstable}, showing that TGF$\beta_1$ provides most
ASI about time of relapse, followed by VCAM1, IL6sR, and uPA, which
together with age and PSA give about 88\% of the available ASI.

\begin{table}[ht]
\begin{center}
\begin{tabular}{rrrr}
& \multicolumn{3}{c}{\textbf{(nats)}}\\
\textbf{Variables} & \textbf{2.5\% ASI} & \textbf{mean ASI} & \textbf{97.5\% ASI}\\
Age and PSA only     & 0.015 & 0.047 & 0.078 \\
+ also TGF$\beta_1$  & 0.071 & 0.117 & 0.163 \\
+ also VCAM1         & 0.129 & 0.181 & 0.234 \\ 
+ also IL6sR         & 0.149 & 0.203 & 0.257 \\
+ also uPA           & 0.151 & 0.203 & 0.257 \\
+ rest               & 0.180 & 0.232 & 0.290 \\
\end{tabular}
\caption{ASI provided by a greedy incremental approach to adding
  biomarkers to the array of explanatory variables.
\label{biomarkerstable}}
\end{center}
\end{table}

\clearpage

\section{Discussion}
\label{discussion}

This model is as far as we know only the second to report
statistically significantly positive ASI about relapse time after
prostatectomy in patients with prostate cancer. We believe it is a
significant advance that approximately double the ASI has been
obtained than using the previous log-skew-Student model. The authors
are aware of other as yet unpublished applications of this model to
other situations where predictions of time of survival from potential
explanatory variables are of interest, where also statistically
significantly positive ASI is obtained. We believe that many other
future applications could exist.

At least one of the authors was very surprised to find that the
information (ASI) provided by this array of biomarkers about future
relapse time was so much greater than that provided by pre- and
post-op Gleason grades, MRI-discoverable variables, and surgical
findings combined. Nonetheless, this becomes perhaps slightly less
surprising in consideration of the recent commercial announcement in
the media\cite{EDXnews} that measurement of an array of over one
hundred biomarkers in blood and urine can accurately determine the
presence or absence of prostate cancer -- though at the time of
writing there are no publicly available details on exactly what is
meant by ``accuracy'' in this announcement or exactly what biomarkers
are being measured.

Of course, while information on relapse time after prostatectomy is
interesting, it would be even more interesting to see how much
information is available about the future time course of possible
prostate cancer that has not been treated and is instead being managed
by watchful waiting. Datasets on such patients exist
(e.g.\cite{ProtecTHamdy}), but as far as we know do not contain
information on biomarkers. Nonetheless it would even be interesting to
know whether in the watchful waiting situation prostate biopsy and
Gleason grading would provide more ASI about time course than it does
in a population who have all had radical prostatectomy.

\textbf{Author Contributions}: Tommy Walker Mackay wrote the new parts
of the modelling software and designed the resampling and annealing
schemes; Mingtong Xu wrote the software for modelling the distribution
of individual ASI samples and inferring their mean; Roger Sewell
proposed the model, executed the analysis, and wrote the paper; and
Shahrokh Shariat collected the original dataset.

\textbf{Funding}: Tommy Walker Mackay and Mingtong Xu thank Trinity
College Cambridge (Woods fund) for support. Mingtong Xu was also
supported by a Wellcome Trust Principal Research Fellowship
(223103/Z/21/Z) and National Institutes of Health grant (5 R01
AI054503) to Lalita Ramakrishnan.

\textbf{Open Access}: For the purpose of open access, the author has
applied a CC BY public copyright license to any Author Accepted
Manuscript version arising from this submission.  This work is
licensed under a Creative Commons Attribution 4.0 International
License.

\newpage

\bibliography{ms}
\bibliographystyle{ieeetr}

\end{document}